\documentstyle[12pt]{article}
\begin{document}
\begin{center}
\large{\bf Some Exact Solutions For The Classical Hall
Effect In Inhomogeneous Magnetic Field}\\

\medskip{\footnotesize\bf\it A.~V.~Chaplik}\\ \footnotesize\it
Institute of Semiconductor Physics, Novosibirsk, 630090, Russia
\end{center}

\begin{minipage}[t]{140mm}
\footnotesize

~~~The classical Hall effect in inhomogeneous systems is
considered for the case of one-dimensional inhomogeneity. For a
certain geometry of the problem and for the magnetic field
linearly depending on the coordinate the density of current
distribution corresponds to the skin-effect.  \end{minipage}
\vspace{1mm}
PACS\ \ 72.15.Gd \\
\vspace{3mm} \normalsize

\section*{Introduction}
     The behaviour of $2D$ electrons in a spatially nonuniform
     magnetic field is of interest in various aspects. The
     simplest situation is one-dimensional inhomogeneity when the
     normal component of the magnetic field $B_z$ varies only in
     one direction, say $B_z(y)$. M$\ddot{u}$ller \cite{1} has
     considered the ballistic regime of a $2D$ electronic system
     for a special case $B_z(y)=B_0ky$ and for the geometry when
the current flows in the $x$-direction perpendicularly to the
     magnetic field gradient. Numerical solution of the
     Schr$\ddot{o}$dinger equation carried out in \cite{1} gives
     the current distribution in the $y$-direction $j_x(y)$.
             Nonzero values of $j_x(y)$ (without electric field) arise
     because the Landau degeneracy is lifted in the inhomogeneous
     magnetic field and the eigenstates become current-carrying.
     Of course, the total current $J=\int j_x(y)dy$ equals zero
     in the absence of the electric field (the case with
     electric field has not been considered in \cite{1}).

     In the present paper $I$ will consider the classical
     magnetotransport for which the local relation between the
     current density $\vec{j}$ and the electric field $\vec{E}$
     is valid
     \begin{equation}\label{1}
     j_\alpha=\sigma_{\alpha\beta}(y)E_{\beta},
     \end{equation}
     where $\alpha$,$\beta$ label the Cartesian components of
     the magnetoconductivity tensor $\hat{\sigma}$. The
     inhomogeneity is supposed to be one-dimensional (in
     $y$-direction) but both possible geometries of the
     experiments ($\vec{j}\parallel Oy$ and $\vec{j}\perp$ Oy)
     will be considered. In the geometry of ref. \cite{1}
     $\vec{j}\perp Oy$ the exact analytical (and quite simple)
     solution is possible for an arbitrary dependence
     $\hat{\sigma}(y)$, including the case of nonuniform
     magnetic field. In the other geometry $\vec{j}\parallel Oy$
     the exact analytical (and rather unexpected) result is
     obtained for the M$\ddot{u}$ller model $B_z=B_0ky$: there
     exists specific static skin-effect when the current density
     exponentially depends on the transversal coordinate $x$.

     In both cases $I$ will consider a specimen in the form of a
     strip of finite width and infinite length. The total
     current $J$ is fixed (measured by ampermeter) while the
     electric field components $E_x$ and $E_y$ are to be found
     from the equations (\ref{1}) and
     \begin{equation}\label{2}
     div\vec{j}=0,~~rot\vec{E}=0.
     \end{equation}
     In such a positioning of the problem one does not need to
     solve the Poisson equation.
\section*{Current perpendicular to the direction of
     inhomogeneity}
     The system of equations (\ref{1}) and (\ref{2}) written in
     the components reads:
     \begin{equation}\label{3}
     j_x=\sigma_0 E_x+\sigma_1 E_y,~~j_y=-\sigma_1 E_x+\sigma_0
     E_y,~~\frac{\partial j_x}{\partial x}+\frac{\partial
     j_y}{\partial y}=0;~~\frac{\partial E_x}{\partial
     y}-\frac{\partial E_y}{\partial x}=0,
     \end{equation}
     where $\sigma_0=\sigma_{xx}=\sigma_{yy}$,
     $\sigma_1=\sigma_{xy}$.

     Look for the solution with $j_y\equiv 0$. Strictly speaking
     one needs the solution with $j_y(y=0)=j_y(y=L)=0$, where L
     is the width of the strip. However due to the evident
     unicity of the solution the assumption made does not
     violate the generality. Then we have:
     \begin{equation}\label{4}
E_y=\frac{\sigma_1}{\sigma_0}E_x=k(y)E_x,
~~j_x=\bigg(\sigma_0+\frac{\sigma_1^2}
{\sigma_0}\bigg)E_x=q(y)E_x,
~~k\equiv\frac{\sigma_1}{\sigma_0},~~q\equiv\frac{
     \sigma_0^2+\sigma_1^2}{\sigma_0}=\frac{1}{\rho_0},
     \end{equation}
     where $\rho_{\alpha\beta}$ is the magnetoresistance
     tensor, $\rho_0=\rho_{xx}$. Further:
     \begin{equation}\label{5}
     div\bar{j}=\frac{\partial j_x}{\partial
     x}=q(y)\frac{\partial E_x}{\partial
     x}=-q(y)\frac{\partial^2\Phi}{\partial x^2}=0
     \end{equation}
     with $\vec{E}=-\vec{\nabla}\Phi(x,y)$.

     The general solution of eq. (\ref{5}) has a form
     \begin{equation}\label{6}
     \Phi=A(y)x+B(y).
     \end{equation}
     It follows from eq. (\ref{4})
     \begin{equation}\label{7}
     \frac{\partial A}{\partial y}x+\frac{\partial B}
     {\partial y}=k(y)A(y).
     \end{equation}
     Hence,
     \begin{equation}\label{8}
     A=const,~~B=A\int\limits_{y_0}^{y}k(y')dy'
     \end{equation}
     The fixed total current determines the value A:
     \begin{equation}\label{9}
     J=\int\limits_{0}^{L}j_xdy=-A\int\limits_{0}^{L}q(y')dy',
     \end{equation}
     and the problem is solved. The Hall voltage defined as
     $\Phi(x,0)-\Phi(x,L)$ equals
     \begin{equation}\label{10}
     V_H=A\int\limits_{0}^{L}k(y')dy'=
     -\frac{J\int\limits_{0}^{L}k(y')dy'}
     {\int\limits_{0}^{L}q(y')dy'}
     \end{equation}
     and one gets the following expression for the {\it
     effective} Hall resistance
     \begin{equation}\label{11}
     \rho_{1}^{eff}=\frac{\langle\frac{\rho_1}
{\rho_0}\rangle}{\langle
     \frac{1}{\rho_0}\rangle},
     \end{equation}
     where the brackets mean averaging over y:
     \begin{equation}\label{12}
     \langle u\rangle\equiv\frac{1}{L}\int\limits_{0}^{L}u(y)dy.
     \end{equation}
     The diagonal component of $\rho_{\alpha\beta}^{eff}$ is
     \begin{equation}\label{13}
     \rho_{0}^{eff}=\bigg\langle\frac{1}{\rho_0}\bigg\rangle^{-1},
     \end{equation}
     that simply corresponds to the parallel connection of
     conducting filaments stretched along the current direction.

     The eqs. (\ref{11}) and (\ref{13}) are valid for any kind
     of a one-dimensional inhomogeneity (e.g. carrier
     concentration, magnetic field, the density of scatterers).
     If the specimen is homogeneous and only magnetic field
     depends on $y$ the relations hold between $\sigma_0$ and
     $\sigma_1$:
     \begin{equation}\label{14}
     \frac{\sigma_1}{\sigma_0}=\lambda B(y),~~\lambda=const,~~
     \frac{\sigma_0^2+\sigma_1^2}{\sigma_0}\equiv\sigma=const,
     \end{equation}
     following from the classical Drude kinetic theory; $\sigma$
     is the Drude conductivity for $B=0$. Then
     $\rho_1^{eff}=\langle\rho_1\rangle$ that corresponds to the
     sequential connection of the Hall voltages created by the
     magnetic field in each conducting filament parallel to
     $Ox$. Thus, in the geometry considered in this section,
     for a homogeneous specimen and inhomogeneous magnetic field
     the results are quite trivial:  \begin{equation}\label{15}
     \rho_0^{eff}=\bigg\langle\frac{1}{\rho_0}\bigg\rangle^{-1},~~
     \rho_1^{eff}=\langle\rho_1\rangle.
     \end{equation}
     Note that for an inhomogeneous specimen, instead of the
     second relation of eq. (\ref{15}), a more complicate result
     (\ref{11}) is valid.

\section*{Current parallel to the magnetic field gradient.}

     Consider now a strip parallel to $Oy$ and look for the
     solution with $j_x(y)\equiv 0$. The specimen is supposed to
     be homogeneous, the local values $\sigma_0(y)$ and
     $\sigma_1(y)$ are determined by the classical kinetic
     theory:
     \begin{equation}\label{16}
     \sigma_0=\frac{\sigma}{1+(\lambda
     B)^2},~~\sigma_1=\frac{\sigma\lambda B}{1+(\lambda
     B)^2},~~B=B(y).
     \end{equation}
     Then from eqs. (\ref{3}) with $j_x=0$ one can easily
     obtain:
     \begin{equation}\label{17}
     j_y=\sigma E_y,~~\frac{\partial^2\Phi}{\partial y^2}=0,~~
     E_x=-\lambda B(y)E_y,
     \end{equation}
     where the last relation in eq. (\ref{17}) follows again
     from $j_x=0$. Hence,
     \begin{equation}\label{18}
     \Phi=C_1(x)y+C_2(x),~~\frac{\partial C_1}{\partial x}y +
     \frac{\partial C_2}{\partial x}=-\lambda B(y)C_1(x),
     \end{equation}
     and for the M$\ddot{u}$ller model $B(y)=B_0ky$ the
     solution has a form
     \begin{equation}\label{19}
     C_1=Ce^{-\lambda B_0kx},~~C_2\equiv 0,~~\Phi=Cye^{-\lambda
     kxB_0}.
     \end{equation}
     From eq. (\ref{19}) one obtains
     \begin{equation}\label{20}
     E_x=C\lambda B_0kye^{-\lambda B_0kx},~~E_y=-Ce^{-\lambda
     B_0kx},~~j_y=-\sigma Ce^{-\lambda B_0kx},
     \end{equation}
     where the constant $C$ can be found via the total current:
     \begin{equation}\label{21}
     J=\int\limits_{0}^{L}j_ydx=-\frac{\sigma C}{\lambda
     B_0k}(1-e^{-\lambda B_0kL}).
     \end{equation}
     The Hall voltage:
     \begin{equation}\label{22}
     \Phi(x=0, y)-\Phi(x=L, y)=y\frac{J\lambda B_0k}{\sigma},
     \end{equation}
     and the Hall resistance in the point $y$ is
     \begin{equation}\label{23}
     \rho_{H}^{(y)}=\frac{\lambda
     B_0ky}{\sigma}=\rho\omega_c(y)\tau,
     \end{equation}
     where $\rho=1/\sigma$, $\tau$ is relaxation time and
     $\omega_c(y)$ is the local value of the cyclotron
     frequency: $\omega_c(y)\tau=\lambda B(y)$. The most
     remarkable feature of the obtained solution is the
     exponential distribution of the current  density along the
     $x$-direction (see eq.  (\ref{20})). This can be called the
     static skin-effect.  Depending on the signs of $J$, $B_0$
     and $k$ the current in $y$-direction is concentrated either
     at the left ($x=0$) or at the right (x=L) edge of the
     strip.  The depth of the skin layer $l_s$ is defined by the
     magnetic field gradient:  $l_s=1/k\omega_{c0}\tau$, where
     $\omega_{c0}$ is the cyclotron frequency for $B=B_0$. The
     electric field depends also exponentially on the
     transversal coordinate. Hence, when measuring the Hall
     voltage $V_H$ between the left edge of the strip $x=0$ and
     some variable point $x$ (for the same $y$!) inside the
     strip one will find the exponential dependence $V_H(x)$ that
     would be an experimental evidence of the skin-effect.
\section*{Alternate electric field.}
     The above obtained results can be easily extended to the
     case of finite frequency $\omega$ of the electric field if
     $\omega <<1/\tau_M$, where $\tau_M$ is the Maxwell
     relaxation time. For the $3D$ situation
     $1/\tau_M=4\pi\sigma_{3D}$ and for $2D$
     $1/\tau_M=2\pi\sigma_{2D}/L$. The parameter $\omega\tau$
     can be of an arbitrary magnitude. By making use of the well
     known formulae for $\sigma_{\alpha\beta}(B)$ allowing for
     the dispersion one can easily see that it is necessary just
     to substitute
     \[\sigma\rightarrow\frac{\sigma}{1-i\omega\tau}\] in all
     the preceeding formulae. For example, the Hall voltage
     between the points (o, y) and (x, y) reads
     \begin{equation}\label{24}
     V_H(o, x; y)=CyR_e\bigg\{\bigg[1-exp\frac{(-kx)\omega_{c0}
     \tau(1+i\omega\tau)}{1+\omega^2\tau^2}\bigg]e^{-i\omega
     t}\bigg\}.
     \end{equation}
     Thus the voltage and the current density decay with
     oscillations when the distance from the strip
     edge increases.
\section*{How to realize the linearly nonuniform magnetic
field.}

     Here I consider only the case of a $2D$ system. As $2D$
     electrons "feel" only the normal component of the magnetic
     field $B_n$ the inhomogeneity of $B_n$ can be achieved for
     a thin conducting film bent to a proper shape and placed in
     a uniform field $\vec{B}(o, o, B)$. The dependence
     $B_n(y)=B_0ky$ is realized for the cylindrical surface
     $z=F(y)$, where
     \begin{equation}\label{25}
F(y)=\pm\frac{1}{2k}\bigg[\sqrt{2ky(1-2ky)}+\arcsin\sqrt{2ky}\bigg],
     ~~ 0\leq ky\leq\frac{1}{2}.
     \end{equation}

     In conclusion, the exact analytical solutions are obtained
     for the classical Hall effect in inhomogeneous systems. If
     the current flows perpendicularly to the inhomogeneity the
     solution is possible in the quite general form. For the
     parallel orientation of the current and magnetic field
     gradient the analytical solution is found for the linearly
     inhomogeneous magnetic field. In the latter case the static
     skin-effect occurs for a specimen in the form of a long
     strip.

     This work has been supported by the Russian Foundation for
     Basic Researches (grant 99-02-17127) and by NWO.

\begin{center}\rule{50mm}{.4pt}\end{center}
\renewcommand{\refname}{}

\end{document}